\documentclass[a4paper,11pt]{article}
\usepackage{pos}
\usepackage{hyperref}

\title{Star clusters in the gamma-ray sky}

\author*[a]{Giada Peron}

\affiliation[a]{INAF, Osservatorio Astrofisico di Arcetri \\
  Largo Enrico Fermi 5, Florence ,Italy}

\emailAdd{giada.peron@inaf.it}

\abstract{Massive Star Clusters (SCs) have been proposed as additional contributors to Galactic Cosmic rays (CRs), to overcome the limitations of supernova remnants (SNRs) to reach the highest energy end of the CR spectrum. Thanks to fast mass losses due to the collective stellar winds, the environment around SCs is potentially suitable for particle acceleration up to PeV energies, and their energetics is enough to account for a non-negligible fraction of the Galactic CRs. Anyhow, the theoretical expectations need to be corroborated by clear observations. Despite the increasing number of detections at different energies, the contamination of other sources often makes it difficult to constrain the contribution arising from stellar winds only, unless one selects objects younger than a few million years, namely before stars start to explode inside clusters.
I will review the results obtained with Fermi-LAT data towards a few massive young star clusters and discuss what implications these result have, especially concerning their contribution to the bulk of Galactic CRs.
}

\FullConference{
}

\begin{document}
\maketitle

\section{Why discussing star clusters at a cosmic-ray conference?}
\begin{figure}
    \centering
    \includegraphics[width=0.65\linewidth]{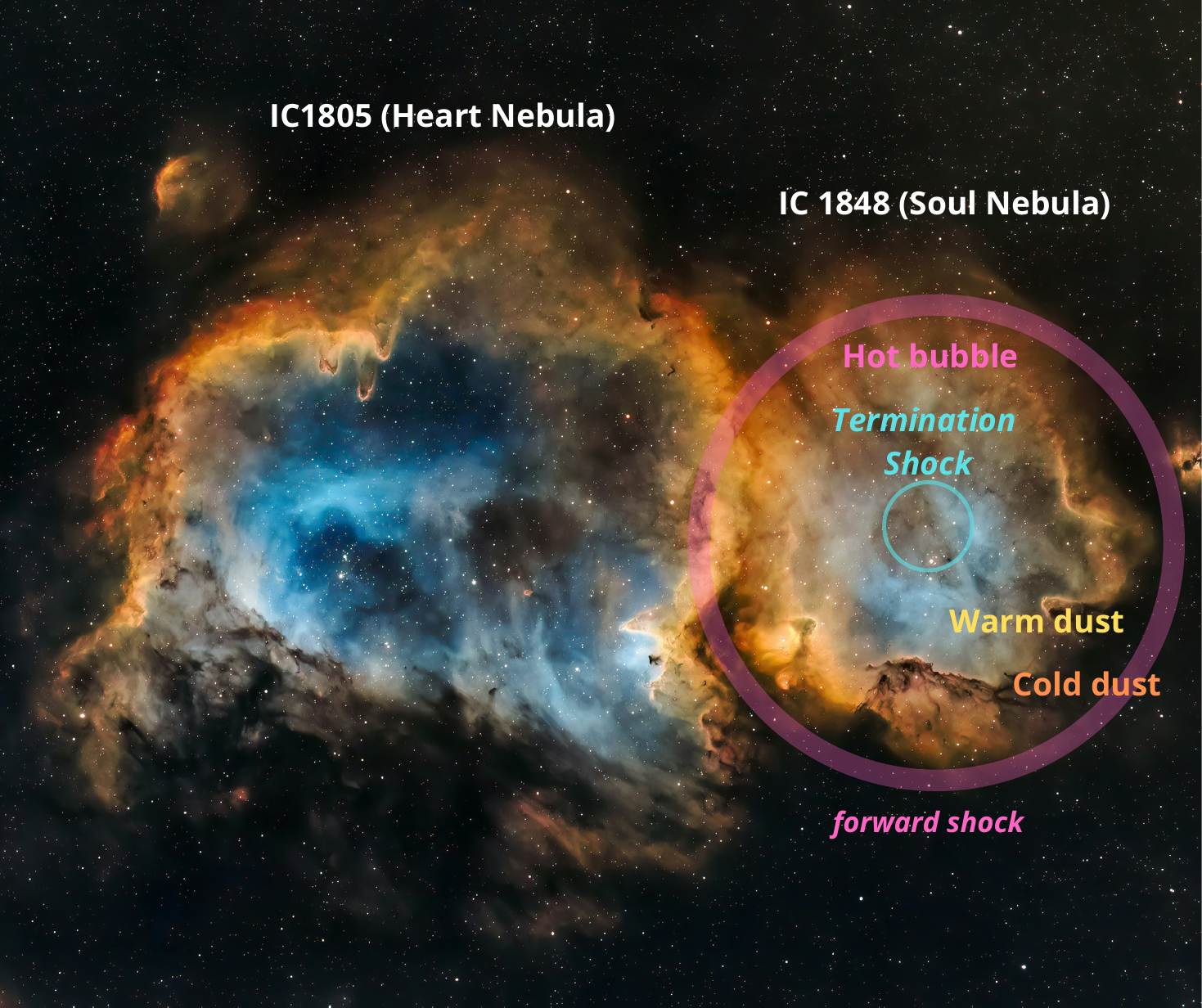}
    \caption{The regions around the Heart and Soul nebulae, namely the H\textsc{ii} regions blown by the embedded clusters IC~1805 and IC~1848, respectively. For an illustrative purpose, the expected size of the termination shock and of the wind-blown bubble are shown. The action of the wind creates typical filaments known as “fingers” visible the surrounding dust. The thick layer of gas and dust around the bubble serves as target for gamma-ray production [Background image by \href{https://unsplash.com/it/foto/un-primo-piano-di-una-formazione-stellare-nel-cielo-_4a3ROTYL5k}{Arnaud Giraud, retrieved from Unplash}]}
    \label{fig:cluster_bello}
\end{figure}

Soon after the release of pioneering data from the Galactic plane by COS-B \citep[][]{Montmerle1979, Cesarsky1983}, it emerged that Galactic gamma-ray sources, and consequently cosmic-ray (CR) sources must be linked to massive stars. Massive stars are shortly lived and most of them ($\sim$ 70-80\% \citep{Oey2004}) are found in groups, referred to as star clusters (SCs) if dynamically bound, or associations if dynamically unbound. Therefore core-collapse supernova (SN) explosions mostly happen within the group of stars in which they formed. As SN remnants (SNRs) have been confirmed to be efficient particle accelerators, the correlation of gamma-ray sources with stellar-rich environments was soon explained. On the other hand, various observations later challenged the hypothesis that CRs are accelerated in SNRs alone, unveiling the need of finding other accelerators within the Galaxy.  The main arguments supporting this idea are two: i) no SNR has displayed clear observational evidence of acceleration of CRs beyond a few tens of TeV \citep[][]{Funk2015, Cao2025}, while the limit for Galactic protons is notably believed to be 3 PeV \citep[see e.g.][]{Gabici2019,AmatoCasanova}; ii) the CR composition deviates from the observations near the solar system and in particular, an excess in the isotopic ratio $^{22}$Ne/$^{20}$Ne is recorded, difficult to explain considering the CR composition in SNRs, but naturally explained assuming that a fraction of GCRs are accelerated from the wind material of massive stars \citep[see extended discussion in][]{Tatischeff2021, Gabici2024}. These two arguments lead the community to reconsider alternative classes of sources as CR accelerators and, in particular, to consider winds of massive SCs. A massive $\sim30~ \mathrm{M}_{\odot}$ star is characterized by a mass-loss rate of $\sim 10^{-6} ~ \mathrm{M}_{\odot}~\mathrm{yr^{-1}}$ , at a velocity of 1000-3000 km s$^{-1}$, for its entire main sequence phase. The wind speeds up, reaching up to 5000 km s$^{-1}$ \citep{Sander2019} in the so-called Wolf-Rayet (WR) phase, which stars of masses $\gtrsim 20~\mathrm{M}_{\odot}$ undergo for $\sim10^5$ yr . The luminosity of main sequence massive stars are on average about $10^{36}$ erg s$^{-1}$, and it is enhanced to  $10^{38}$ erg s$^{-1}$ in the WR stage. Considering that there are about 10$^5$ OB stars, and a thousand Wolf-Rayet stars are inferred in the Galaxy \citep[][]{Crowther2015}, one expects a total power of $1-2\times 10^{41}$ erg s$^{-1}$ contributed by stellar winds \citep{Seo2018}. The latter is approximately 10\% of the power contributed by SNRs and is comparable with the power of Galactic CRs, \citep[$\sim 7 \times 10^{40}$erg~s$^{-1}$;][]{Strong2010}, making stellar winds promising CR contributors in the Galaxy, if a fraction of this power is eventually converted into accelerated particles. A collective wind, resulting as the superposition of single stellar winds may form if the SC is compact enough \citep{CantoRaga2000}. The resulting collective wind, is typically more dense and hence more powerful as $L_{w} \sim 0.5 \sum_{i} \dot{M}_i <v>^{2}$, where $\dot{M}_i$ is the mass-loss rate of each star, and $<v>$ is the average wind velocity.    
Theory of diffusive shock acceleration at the collective cluster wind termination shock was developed and extensively described in \cite{Morlino2021}. Compared to SNRs, The advantages of developing shock acceleration in SCs are twofold: firstly, the shock endures for a longer time ($\mathcal{O}(10^6~\mathrm{yr})$) and with a constant velocity, different from the shock of SNRs, which rapidly slows down; secondly, SC's termination shocks are surrounded by a hot turbulent bubble (see schematic geometry in Fig \ref{fig:cluster_bello}) that guarantees longer confinement of accelerated particles, therefore making the acceleration much more effective. 

These premises make SCs interesting potential accelerators, which under certain assumptions on luminosity and level of turbulence could reach maximum energies of the order of PeV. Beyond the quest of Galactic hadronic PeVatrons, if confirmed that stellar clusters contribute a small fraction of Galactic CRs, as required to explain the CR composition anomalies \citep[][]{Tatischeff2021}, they should be considered in the description of the Galactic ecosystem, especially with their natural connection with regions where stars are still forming, being CRs one of the major sources of feedback for the formation process. Moreover, having a second contributor to Galactic CRs, beyond SNRs, may help to understand the features that are observed in the CR spectrum measured at Earth. 

\section{What makes a star cluster a gamma-ray source?}
Gamma-ray observations are key to investigate the properties of SCs as Galactic accelerators. If favorable environmental conditions apply, such as large gas densities or enhanced radiation densities, and the acceleration efficiency is high enough, SCs are expected to be bright gamma-ray sources. In Table \ref{tab:detected_scs}, one can see a list of SCs with a gamma-ray counterpart either in the GeV or in the TeV energy range. When compared to the number of massive SCs known in the Galaxy, $\sim$ 7000 \cite{Hunt2023}, the number of known gamma-ray emitting SCs is very small, suggesting that environmental conditions may hinder the detection. These sources are found usually to be extended, another challenge for recognizing SCs in the gamma-ray sky both due to the enhanced background \citep[see e.g.][]{CelliPeron2024} and to the source confusion. Still, if these sources are efficient emitter but just below threshold of our current detectors, they may contribute to the Galactic diffuse emission. Interestingly, a contribution of a population of SCs that convert 10\% of their energy in accelerated particles could explain the excess measured by LHAASO in the diffuse emission up to tens of TeV \cite{Menchiari2025}.

While it is difficult to make a census of detected SCs given the very small number of cases, a few considerations can be put forward. For example, from the list in Table \ref{tab:detected_scs}, it seems that TeV detection is favored in the case of more evolved system, where there is at least one WR star, while Femi-LAT seems able to reveal also very young systems, where stars are supposedly still in the main sequence. This could hint to the need of WR stars to push the spectrum to very high energy, but only further observations could confirm this. Moreover, we see that the number of very young (<3 Myr) star clusters detected in gamma rays generally exceeds the number of older systems. A reason for this could be that very young stellar clusters are still surrounded by the dense gas out of which the star formed, that serves as a target to enhance the emission. Older SCs have, instead, already depleted and swept away all the gas that surrounds them. Meanwhile, the size of the emitting region will be larger for older clusters, as both the wind termination shock and the wind-blown bubbles get larger with time, making their surface brightness dimmer. In this regard, Westerlund~1 may be one of the few lucky cases (Westerlund~1 is the most massive SC in the Galaxy) where the emission at the termination shock is visible. In the lack of gas as a target, in fact, the emission is interpreted as inverse Compton scattering over the intense UV radiation field provided by the stars \cite{Harer2023}.  

A similar trend of favoring young embedded SCs for gamma-ray detection is also confirmed by the cross-correlation carried out in \cite{Peron2024b}. In that work, catalogs of known SCs, traced either by their stellar light \citep[as seen by Gaia][]{Cantat-Gaudin2020,Hunt2023} or by their infrared emission induced on their surrounding dust (for simplicity referred to as H\textsc{ii} regions) \citep{Anderson2015} were compared. The star clusters traced by Gaia are in general quite older than the one traced by infrared emission: as a consequence of an observational bias, very young star clusters, being embedded in the gas cocoon out of which they formed, are often invisible in optical light and must be searched as infrared-bright H\textsc{ii} regions. The analysis unveiled a very high degree of correlation between H\textsc{ii} regions, and unidentified gamma-ray sources from the Fermi-LAT catalog. On the other hand, no significant correlation emerges when considering Gaia-identified star clusters, probably due to the larger size and to their more diluted circumstellar gas. This result suggests that H\textsc{II} regions are good tracers for locating gamma-ray emitting star clusters. Note that, besides Westerlund~1, whose GeV emission is displaced from the location of the cluster, the other firmly detected clusters listed in the upper part of Table \ref{tab:detected_scs} all emerge in the cross-correlation between HII regions and Fermi-LAT sources. 

\begin{table}
\small
    \centering
    \begin{tabular}{l|cccc}
         Name &Age   &N$_{\mathrm{WR}}$&  GeV detection & TeV detection \\
         \hline \hline
         Westerlund~1 & 4 Myr &24&  Fermi \citep{WesterlundOhm} (Outflow \citep{outflow2025}) &  H.E.S.S. (Ring) \citep{Aharonian2021}\\
         Westerlund~2 & 2--3 Myr &2&  Fermi \citep{Mestre2021Probing2} &  H.E.S.S. \citep{AharonianWd2} \\
         Cygnus OB2 & 2--3 Myr &3&  Fermi\citep{CygnusFermi,Cygnus2023}&  HAWC \citep{Abeysekara2021}, LHAASO \cite{Cao2021a}\\
         NGC~3603 &  2--3 Myr &5&  Fermi \citep{Yang2017,Saha2020}&  --\\ 
  RCW~38 & $\lesssim$ 1 Myr&--& Fermi\citep{Peron2024a}&--\\
 RCW~36& $\sim$ 1 Myr &--& Fermi \citep{Peron2024a}&--\\
 RCW~32 & $\lesssim 2$ Myr &--& Fermi \citep{Peron2024a}&--\\
 NGC~6611 (M16)& 1.3~Myr &--& Fermi \citep{Peron2025}&--\\
 NGC~6618 (M17)& 1 Myr & --& Fermi \citep{LiuM17} &--\\
 W40 (RCW~174)& $<$ 1 Myr & -- & Fermi \citep{w40}&--\\
 Danks~1& 1.5 Myr &6& Fermi \citep{danks}&--\\
 Danks~2& 3 Myr  &1& Fermi \citep{danks}& --\\
 \hline
W43 complex & &1& Fermi \citep{Yang2020w43} & H.E.S.S.,HAWC,LHAASO (PWN) \\
$[$BDS 2003$]$~8 &  1~Myr  &--& -- & HAWC \cite{Albert2021} (PWN)\\
RSGC~1 & 10--14 Myr   &--& Fermi \citep{sun2020} & H.E.S.S., MAGIC (SNR,PWN) \\
Mc20 & 3--8 Myr &--& Fermi\citep{FermiMc20} (SNR,PWN) & H.E.S.S. (PWN)\\
W33 (Cl 1813) & 4 Myr & 2 &  Fermi \citep{Guo2024} (SNR,PWN)&  H.E.S.S. (PWN)\\
Mercer~81 &  4 Myr & 8 & --& H.E.S.S. \citep[][]{davies2012}(SNR)\\
SGR1806-20 & 3--4 Myr & 4 & Fermi \citep{Yeung2016} (SNR) & H.E.S.S.  (Magnetar)\cite{hess1806}  \\

    \end{tabular}
    \caption{Known Galactic young star clusters or stellar association detected in gamma rays, and the instrument that allowed the detection (for compactness we shorten the name "Fermi-LAT" in "Fermi"). In the upper part of the table, we report cases where the identification of the emission is likely, as result of a deep study, or because there are no alternative counterparts. In the lower part of the table we report tentative associations, namely those cases where the star cluster is only one of the possible counterparts of the gamma-ray emission; we indicate possible other identification in parenthesis. Where no specific reference is reported we refer to the main catalogs of the instruments, that can be retrieved at \href{https://www.tevcat.org/}{https://www.tevcat.org/}. }
    \label{tab:detected_scs}
\end{table}
\section{Gamma-ray observations of star clusters: what do they tell us?}
\begin{figure}
    \centering
    \includegraphics[width=0.33 \linewidth]{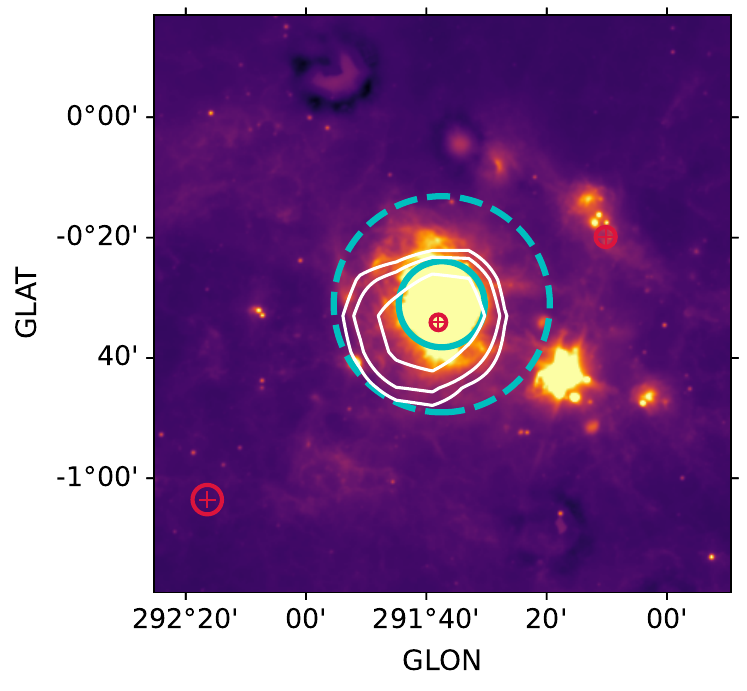}\includegraphics[width=0.66\linewidth]{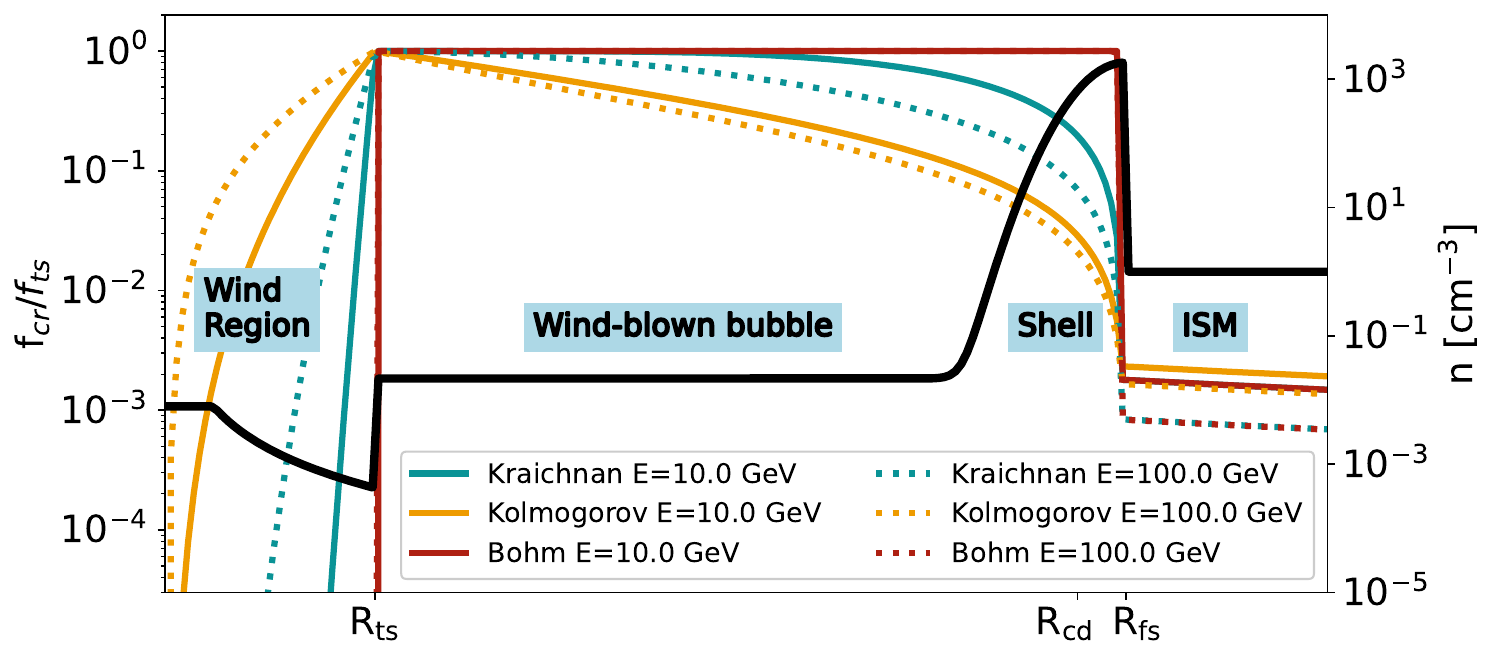}
    \caption{On the left: the H\textsc{ii} region surrounding NGC~3603 as seen by WISE at 22$\mu$m; the size of the wind-blown bubble calculated using Weaver formula and $\eta_M=100\%$ or $\eta_M=1\%$ are shown as a dashed and a solid turquoise circle, respectively. The position of unidentified Fermi-LAT sources are indicated as red circles with an inner cross. The contours indicate significance detection levels obtained with Fermi-LAT observations at 9, 10 and 12$\sigma$ as reported in \citep{Peron2025Ricap}. On the right: the radial distribution of the gas and of the CRs for increasing distance from the center of the cluster \citep[Figure adapted from][]{Peron2025}.   }
    \label{fig:hiiregion}
\end{figure}

From detailed analysis of the gamma-ray emission of  a few H\textsc{ii} regions \citep[see details in ][]{Peron2024a,Peron2025} several conclusions emerged. The first result was that, in most cases, the gamma-ray emission closely matches the emission of the H\textsc{ii} regions detected in the infrared domain. We recall that the infrared emission originates at the edge of the hot bubble, where the dust, swept up in a shell, is heated by the radiation of the massive stars (see sketch in \ref{fig:cluster_bello}). In this sense, the size of the H\textsc{ii} region gives an indication of the size of the bubble, an additional indication that H\textsc{ii} regions are good areas where to look for gamma-ray sources. On the other hand, when comparing the size of the bubble calculated according to Weaver's theory \citep[][]{Weaver1977} to the observational size of the considered H\textsc{ii} region, it emerges that the latter is much smaller, suggesting that an important energy loss is taking place. This effect is expected as a consequence of radiative cooling and, according to simulations, the effective fraction of mechanical energy that is used to inflate the bubble, $\eta_m$, ranges from $\sim$0.1\% to $\sim$10\%, being smaller for bubbles surrounded by a denser medium \cite{Yadav2017}. Taking an intermediate value of $\eta_m$ of 1\% one can immediately see how the gamma-ray emission matches the size of the bubble better for the case of the young stellar cluster NGC~3603, whose significance contours, derived from Fermi-LAT observations \citep{Peron2025Ricap}, are shown in the left panel of Figure \ref{fig:hiiregion}, together with the size of the bubble calculated assuming $\eta_m =100\%$ or $\eta_m =1\%$. A correct estimate of $\eta_m$ would be essential for those cases where the stellar wind luminosity is unknown, but other parameters such as the size of the bubble and the gas density are known. Knowing the latter parameters, one can compute the total wind luminosity through:

$$ L_w = \eta^{-1}_m \bigg(\frac{R}{28~\mathrm{pc}}\bigg)^5 \bigg(\frac{n_0}{200~\mathrm{cm}^{-3}}\bigg) \bigg(\frac{\tau_{SC}}{1~\mathrm{Myr}} \bigg)^{-3}  $$ 

and relate it to the gamma-ray luminosity that one derives from observations. One readily example is given by the stellar cluster RCW~38, observed by Fermi-LAT to have an extension of 0.21$^\circ$ \cite{Peron2024a}, similar to the size of its corresponding H\textsc{ii} region (G267.935-01.075), $\sim 5$~pc at the distance of the region. Considering that the system is located inside a dense gas region of $n\sim 2000$~cm$^{-3}$, one obtains $L_w\sim 4\times 10^{36} \mathrm{erg~s^{-1}}(0.01/\eta_m) $, assuming a SC age of 1~Myr. This estimate is compatible with other values found in the literature \citep[][]{Pandey2024}, but it is very sensitive to the age of the cluster, which is not accurately known . Then, assuming that the observed gamma-ray luminosity, $L_{\gamma} = 5 \times 10^{33} \mathrm{erg~s^{-1}}$ is produced in condition of fast cooling \citep[see discussion in][]{Peron2024a}, one can easily calculate the fraction of wind luminosity that is converted into accelerated wind as:
$$ \eta_{CR} = \frac{L_{CR}}{L_w} \approx \frac{3L_\gamma}{L_w} \approx 0.3 \% \bigg(\frac{\eta_m}{0.01}\bigg)  .$$
The latter is a rough estimate of a lower limit of the acceleration efficiency obtained in such a system, since the assumption of fast cooling requires that no particle has escaped the region yet. Allowing some particles to escape would increase $L_{CR}$ and hence $\eta_{CR}$. Having an estimate of $\eta_m$ would, in synthesis, let us obtain an estimate of the CR acceleration efficiency of SCs embedded in H\textsc{ii} regions that would depend only on observational parameters. 

The above mentioned calculation was obtained assuming that the observed gamma-ray emission has a hadronic origin. The similar morphology that emerges between the infrared and the gamma-ray emission suggests that the latter originates in the shell as well, favoring an hadronic scenario indeed, as the gas density peaks in the shell region, as depicted in the right panel of Fig. \ref{fig:hiiregion}. Particles, after being accelerated at the wind termination shock, propagate through the low-density wind-blown bubble and their distribution is shaped in the journey through this turbulent medium and, if the confinement is not efficient enough, the particle may escape from the boundary of the shell before producing  detectable gamma-ray emission. The diffusion coefficient is not known in the bubble, but a magnetic field of  $\sim 10~\mu$G is easily achieved if a fraction of a few percent of the wind luminosity is converted into magnetic energy, and depending on the spectrum of turbulence, the accelerated particles may be distributed differently within the bubble (see right side of Fig. \ref{fig:hiiregion}). A consistent interpretation of the emission should take into account this modulation, as it was done, for example, in the modeling of the emission of the young massive star cluster NGC~6611 \cite{Peron2025}. 

\begin{figure}
    \centering
    \includegraphics[width=0.8\linewidth]{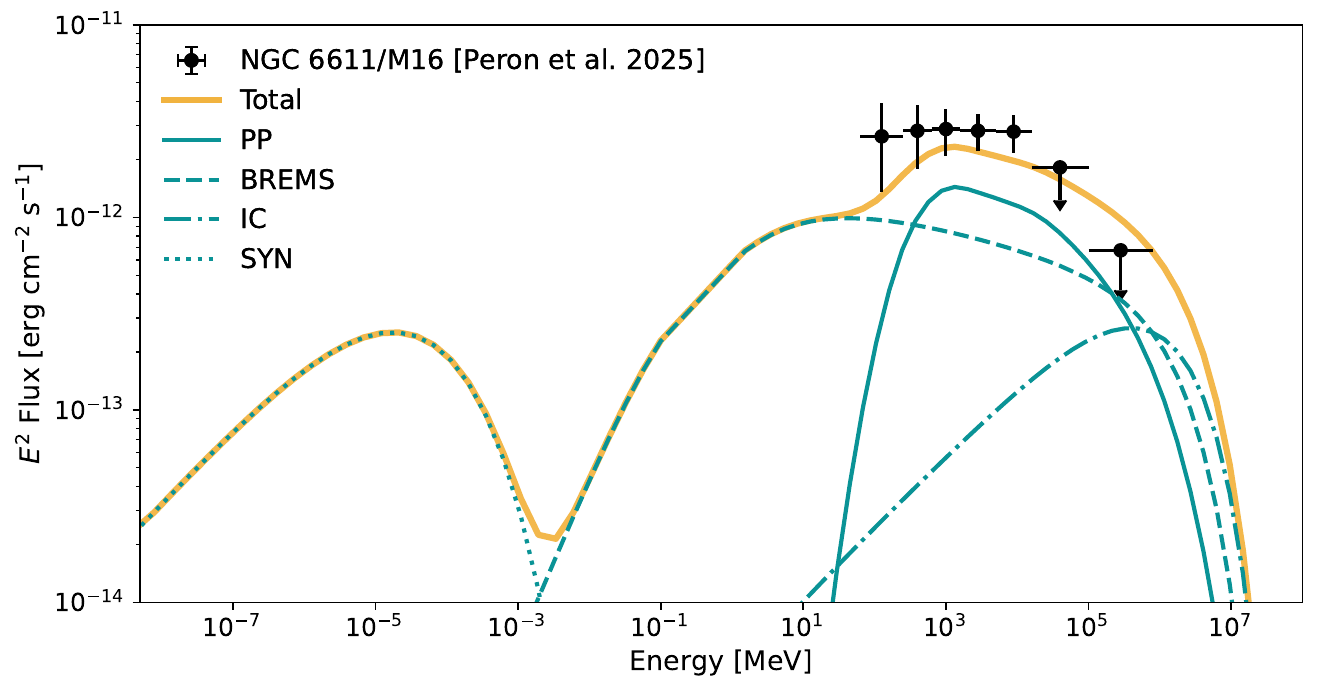}
    \caption{Broad-band emission expected from the SC NGC~6611. The GeV emission is predominantly hadronic, while the leptonic emission is modeled starting from electrons with the same injection spectrum as protons, but assuming a reduced normalization of $K_{ep}=0.015$, and the appropriate energy losses, considering a magnetic field of $\sim 10 \mu G$. \citep[The figure has been adapted from][]{Peron2025}.  }
    \label{fig:broad}
\end{figure}

Formally, the correlation of the emission with a dense gas shell is not an exhaustive argument for claiming a hadronic interpretation, since bremsstrahlung of relativistic electrons over the ions of the gas is also a valid alternative for explaining GeV emission. A leptonic emission, however, could be dominant in a gas-dense region only if $K_{ep}$, namely the ratio of accelerated electrons over protons at the shock, is higher than what usually found in SNRs and in simulations \cite{Peron2024a,Peron2025}, namely $10^{-4}-10^{-3}$ \citep[][]{MorlinoCaprioli2012, Merten2017}. For interpreting the emission as leptonic, one should investigate why the shock in SCs differs so evidently from the behavior of shocks in SNRs. A similar implicit result is also found for Westerlund~1 and Cygnus~OB for which, the interpretation of the VHE energy in terms of leptonic emission requires an electron acceleration efficiency of 0.09-0.28\% for the former \cite{Harer2023} and 1\% for the latter \cite{Haerer2025}. These numbers are formally fine, but assuming a $K_{ep}\lesssim 10^{-3}$ would require an exceedingly large efficiency of protons. This could be reconciled in the case of Cygnus~OB2, assuming that the emission is predominantly hadronic \cite{Abeysekara2021,Menchiari2024}, however, the case of Westerlund~1 is more controversial as this evolved system does not seem to have an appreciable amount of gas in its surroundings, making the hadronic interpretation less convincing. Multi-wavelength observations of radio or X-ray synchrotron could help in constraining the leptonic population and therefore investigating the real nature of this emission. However, a real smoking-gun could come from MeV observations of the region below the pion bump, where one could really distinguish among a bremsstrahlung or a pion origin of the emission and even estimate directly the ratio $K_{ep}$. As one could see for example from the broad-band emission model for the SC NGC~6611 in Fig. \ref{fig:broad}, below a few tens of MeV the emission is naturally dominated by leptonic mechanisms, and therefore measuring simultaneously the MeV and GeV band could provide a direct measure of the ration between hadrons and leptons. 

\section{How much do star clusters contribute to the Galactic cosmic ray population?}
From a dozen SCs, convincing gamma-ray emission is detected in their surroundings, however, a complete investigation of the nature of the emission has been possible only for a few objects. For a clear understanding on whether this source class is an important or a negligible contributor to the Galactic CR pool we must then wait for more gamma-ray observations. Nevertheless, a few considerations can be made starting from the objects that have been studied. For the SCs in the Vela region \citep[RCW~38, RCW~36, RCW~32][]{Peron2024a}, the modeling is made difficult by the lack of a reasonable estimate for the wind luminosity, but using a reasoning similar to what presented in the previous section, and assuming full confinement, a strong lower limit was assessed to be on average $\langle \eta_{CR} \rangle = 0.5$\%. A more detailed modeling was possible for the system NGC~6611 \citep[][]{Peron2025}, for which the wind luminosity could be estimated \citep[e.g][]{Celli2024}, and hence some inference could be done on the magnetic field, assuming that a few percent of the wind kinetic luminosity is converted into magnetic turbulence. This allowed us to write a prescription for the diffusion and advection through the bubble and for the CR distribution accordingly (see, e.g., Fig. \ref{fig:hiiregion}).  The resulting acceleration efficiency depends on the assumptions on propagation, but after all considerations, it is constrained within $\sim$1\% and $\sim$4\%. The estimate concerns protons only, as it is shown that leptonic emission must be subdominant in this system not to violate the derived upper limits \citep[see more details in ][]{Peron2025}.  
The emission of the Cygnus~OB2 association has been analyzed by \cite[][]{Menchiari2024}, assuming that proton acceleration takes place at the collective wind termination shock. In this picture, the acceleration efficiency is constrained within 0.4 and 13\%. Other interpretations of the gamma-ray emission, however, emerged for this object after the release of the data by the LHAASO collaboration \citep[][]{LhaasoCygnus} unveiling emission exceeding a few PeV, and distributed over a huge ($\sim$ 10$^\circ$) area. Both the high-energy spectrum and the morphology are explained by \citep[][]{Haerer2025} assuming that an extraordinary powerful SNR exploded in the bubble $\sim 50$ kyr ago, while they explain the low-energy part of the spectrum as inverse-Compton emission of relativistic electrons accelerated by the stellar wind, assuming that 1\% of the wind power is converted into accelerated particles. Even assuming that the emission is leptonic dominated, in the context of diffusive shock acceleration, we expect a much larger fraction of accelerated protons, therefore we could infer $\eta_{CR}$ >1\%.  A similar reasoning could be put in place for the case of Westerlund~1, whose H.E.S.S.-detected emission \citep{Aharonian2021} is interpreted as inverse-Compton emission at the wind termination shock of the cluster. The derived electron efficiency let us infer $\eta_{CR}> 0.09$ \%. 

In spite of the large differences among these object, in therms of age, number of stars, and spacial distribution, their modeling generally suggests a small acceleration efficiency, almost one order of magnitude smaller than what inferred for SNRs. It should be noted, though, that in most cases the estimated value is a lower limit on the acceleration efficiency of protons. Taking the average of these lower limits, we find $\langle \eta_{CR}\rangle =0.67\%$. With this value in mind, some consideration on the general contribution of winds of star clusters to the overall population of CRs can be carried out. Assuming that the entire population of OB stars emits with such an efficiency,  and recalling that the Galactic population of OB stars contributes with a power of at least $L_{w,tot}\sim 10^{41} \mathrm{erg~s^{-1}}$, we obtain a total power contributed by stellar winds which is $\langle\eta_{CR} \rangle L_{w,tot}\approx 7 \times 10^{38}$ erg~s$^{-1}$ .  Considering an average power of CRs in the Galaxy of $\sim 7\times 10^{40}$  erg~s$^{-1}$ \citep{Strong2010}, this estimate implies that stellar winds contribute at least 1\% of the population of Galactic CRs. We recall that this is a strict lower limit, derived by considering only the lower limits on proton acceleration efficiency, but also a lower limit on the stellar luminosity, which is expected to be enlarged of a factor $\sim$2, when accounting for the luminosity of Galactic WR stars. This value, although small, is approximately enough to account for the anomalies in the chemical composition that would require a fraction of $\sim$5\% of CRs accelerated at stellar winds \citep[see e.g.][]{Tatischeff2021}.

\section{Conclusions and outlook}
Star clusters have been detected in gamma-rays, confirming them as particle accelerators, in different stages of their life: at very-young ages (< 2~Myr) where they are still embedded in their gas cocoon, and where the WR phase is not active yet (e.g. NGC~6611 and RCW~38), at later stages, where they host WRs but still no SNRs (e.g. NGC3603, Westerlund~2, Danks~1 and Danks~2), and also after some SNRs already took place in their environment (e.g. Westerlund~1, Cygnus~OB2).  While in this latter scenario the attribution of the emission is challenged by the contamination by SNRs and their byproducts, such as pulsars and their nebulae, in the earliest phases of their life, the emission must be linked to the stellar wind. The analysis of a handful of objects, reviewed here, unveiled hadronic emission that is interpreted if a small fraction of the wind power is used for particle acceleration. These results point towards $\gtrsim 1\%$ of Galactic cosmic rays accelerated at the wind termination shock of SCs. To confirm this number however, a larger number of system should be analyzed, by putting forward a self-consistent modeling of the propagation through the turbulent bubble and of the emission at the edge of it. Meanwhile, an evaluation of the contribution of isolated stars (including WR stars) should also be investigated, as it is estimated that $\sim$ 20-30\% of massive stars are not part of a group of stars. Measuring their gamma-ray emission however may be challenged by the absence of dense gas: if they exited the dense-gas cocoon out of which SC form, or by the too dim radiation field, that may be an order of magnitude lower for isolated stars compared to SCs or associations \citep{Inventar2025}. At the same time, the possible contribution of binary systems, commonly found within stellar groups, is disregarded in this treatment, but should in future be considered as additional sources of both CRs and gamma-rays giving the promisingly high acceleration efficiency in systems of colliding wind binaries \citep{Bykov2024}.

Meanwhile, an interesting feature emerged connected to Westerlund~1, as claimed by \citep{outflow2025}, an extended outflow unveiled in the GeV band and spatially and spectrally connect to the TeV emission. This indicated both that SCs may be actual drivers of the dynamics of Galactic cosmic rays, but also that reconstructing the full acceleration history of star clusters may, in some cases, require a full and deep investigation of their surroundings.

Finally, while the issue of the CR composition seems to be solved by proving that a small fraction of particles is accelerated at the termination shock of stellar winds, an open question remains on which are the sources able to provide PeV protons at Earth. Very-high-energy emission ($\gtrsim$10 TeV) could be firmly recognized only for three objects. Source confusion and the extension of these SCs make other identifications quite challenging. Despite the difficult interpretation of the origin of the emission, these three objects all show gamma rays up to $\sim$100 TeV, in the case of Westerlund~1 and Westerlund~2, and exceeding PeV, in the case of Cygnus~OB2, making them still promising PeVatron candidates. For understanding whether these are rare examples or they could actually account for the spectrum of CRs around PeV energies, we should wait for the advent of the next generation of gamma-ray observatories, which includes the ASTRI-MiniArray and the Cherenkov Telescope Array Observatory. The latter will improve the sensitivity of one order of magnitude, while improving the angular resolution and the field of view, a paramount feature for observing extended sources. On a theoretical point of view, instead, it was soon noted that to reach PeV energies with winds alone, a very large wind luminosity is needed, together with a great efficiency in confining particles  \citep[][]{Morlino2021,Blasi2023}. Therefore, it remains unclear, at the moment, whether winds of SCs are effectively able to push the maximum energy beyond the SNR limit, by reaching the required PeV regime. Another open scenario remains, where SNRs expand inside a young SCs, where the ambient medium is turbulent enough to allow a more effective magnetic field amplification and, consequently, a larger maximum energy, compared to isolated SNRs.

\paragraph{Acknowledgments} I would like to thank the organizers of the 39th International Cosmic Ray Conference for their invitation as well as the participants for inspiring discussion. I am thankful to G. Morlino for insightful suggestions on how to improve the draft. GP is funded through the INAF Astrophysical fellowship initiative.
\bibliographystyle{abbrvnat}
\bibliography{references}

\end{document}